\documentclass[doublecol]{epl2}

\def\sech{\mathop{\mathrm{sech}}\nolimits}

\title{Interacting dipoles in type-I clathrates: Why glass-like though crystalline?}
\shorttitle{Interacting Dipoles in Type-I Clathrates}

\author{Tsuneyoshi Nakayama\inst{1}\footnote{E-mail: \email{Riken-nakayama@mosk.tytlabs.co.jp}}\footnote{Present address: Toyota Physical and Chemical Research Institute, Nagakute, Aichi 480-1192, Japan.} \and Eiji Kanashita\inst{2}
}
\shortauthor{T. Nakayama \etal}

\institute{
  \inst{1} Materials Science Division, Argonne National Laboratory, 9700 South Cass Avenue, Argonne, IL 60439, USA\\
  \inst{2} Advanced Photon Source, Argonne National laboratory, 9700 South Cass Avenue, Argonne, IL 60439, USA
}

\abstract{
Almost identical thermal properties of type-I clathrate compounds to those of glasses follow naturally from the consideration that off-centered guest ions possess electric dipole moments.
Local fields from neighbor dipoles create many potential minima in the configuration space.
A theoretical analysis based on two-level tunneling states demonstrates that interacting  dipoles are a key to quantitatively explain the glass-like behaviors of low-temperature thermal properties of type-I clathrate compounds with off-centered guest ions.
}

\begin{document}

\maketitle
\section{Introduction}
Clathrate compounds, attracting much attention in connection with thermoelectric materials~\cite{01nolas1}, show unique physical properties at low temperatures.
A series of recent experiments on type-I clathrate compounds,
$\beta$-Ba$_8$Ga$_{16}$X$_{30}$ (X=Sn,Ge) or $\beta$-X$_8$Ga$_{16}$Ge$_{30}$ (X=Eu, Sr, Ba)---hereafter abbreviated by $\beta$-BGX, $\beta$-XGG, or $\beta$-BGS,
\textit{etc}.--- have revealed that the characteristics of thermal~\cite{02avila1,03suekuni1,04avila2,05suekuni2,06suekuni3} and spectroscopic~\cite{07takasu1,08takasu2} properties at low temperatures strongly depend on the states of guest ions in cages.
These studies confirmed earlier measurements on thermal properties for various types of clathrate compounds~\cite{09nolas2,10cohn,11nolas3,12nolas4,13sales,14bentien1,15bentien2}.

The most intriguing findings~\cite{02avila1,03suekuni1,04avila2,05suekuni2,06suekuni3,
07takasu1,08takasu2,09nolas2,10cohn,11nolas3,12nolas4,13sales,14bentien1,15bentien2} are that the specific heats and the thermal conductivities of type-I clathrate compounds with \textit{off-centered} guest ions show almost identical thermal properties  to those of glasses  in the regime below a few K~\cite{16zeller}, in spite of the remarkable difference in microscopic structures among these materials.
The specific heat of $\beta$-BGS is scaled by the relation $C\simeq \alpha T+ \beta T^3$ with $\alpha \simeq 30$ $\mathrm{mJ\,mol^{-1}\,K^{-2}}$ and $\beta\simeq 50$ $\mathrm{mJ\,mol^{-1}\,K^{-4}}$~\cite{02avila1,03suekuni1,04avila2, 05suekuni2,06suekuni3}.
While topological disorder is a key element for glasses~\cite{17nakayama}, the disorder in type-I clathrate compounds is much less relevant, as diffraction studies have provided the evidence that the microscopic structures of type-I clathrate compounds are rather well-defined~\cite{02avila1,03suekuni1,04avila2,05suekuni2, 06suekuni3,07takasu1,11nolas3,13sales}.

A feature found in experiments~\cite{02avila1,03suekuni1,04avila2, 05suekuni2,06suekuni3,07takasu1, 08takasu2,09nolas2,10cohn, 11nolas3,12nolas4,13sales, 14bentien1,15bentien2} is that guest ions take either   an on-center or an off-center position depending on the size of the cages or, equivalently, on the ionic radii of the guest ions.
It is notable that the  thermal properties almost identical to those of glasses are only observed in clathrate compounds with off-centered guest ions.
It remains a mystery why the observations for these clathrate compounds without topological disorder show thermal properties almost identical to those of glasses~\cite{16zeller}.

The experimental data so far have been  analyzed on the basis of the model  that the isolated off-centered guest ion governs the thermal properties at low temperatures.
This simple view often misleads us to the idea that a \textit{noninteracting} guest-ion picture could explain the glass-like behaviors.
Such a \textit{noninteracting} guest-ion picture, however, conflicts with the observation, yielding a specific heat two orders larger than the observed value at 1~K, as demonstrated below.
This Letter provides a clear interpretation of the glass-like specific heat behaviors of the type-I clathrate compounds with off-centered guest ions, $\beta$-BGS~\cite{02avila1,03suekuni1,04avila2, 05suekuni2,06suekuni3}.
Our treatment is general and directly applicable to other types of clathrate compounds with off-centered guest ions.

\section{Off-centered guest ions and tunneling states}
Experimental studies have shown that the positional symmetry of the guest ions in clathrates is \textit{broken}
by increasing the size of the cages, and it has been experimentally confirmed that the guest ions take four  off-center positions at $r_0$ from the center of a cage~\cite{02avila1,03suekuni1,04avila2,05suekuni2, 06suekuni3,07takasu1,11nolas3,13sales}.
A typical distortion $r_0=0.43$ {\AA} in $\beta$-BGS is obtained from diffraction experiments~\cite{02avila1,04avila2}.
This value is about 7.2\% of the distance between the neighboring 14-hedrons ($a/2=5.84$ \AA, where $a$ is the lattice constant).
It is straightforward to see that such an off-centered guest ion induces a large \textit{electric dipole moment} due to the difference of the charges between the Ba$^{2+}$ guest ion and the Ga$^{-}$ ion constituting cages.
The strength of the electric dipole moment becomes $p\approx 4.1$ Debyes in $\beta$-BGS from $r_0=0.43$ {\AA}.
Here, we emphasize that the cages with off-centered guest ions intrinsically possess dipoles and that it is crucial to properly deal with their characters in modeling the systems.

It has been also confirmed~\cite{13sales} that guest ions experience a \textit{hindering} potential $V_h(\theta)$ with a four-fold inversion symmetry along the azimuthal direction.
The barrier height of the hindering potential $V_h$ between nearby potential wells is estimated from the difference between  the largest and the smallest strengths of van der Waals potentials to a guest ion from the framework atoms of a cage, which should be of the order of 10~K.
Actually, first-principles calculations have shown $V_h$ to be $\simeq 20$ K for Sr$^{2+}$ guest ions in $\beta$-SGG~\cite{18madsen}.
By expressing the position of a guest ion in terms of two-dimensional spherical coordinates perpendicular to the four-fold inversion axis centered in a cage, the motion of an off-centered guest ion is mapped onto a rotational motion of dipoles having a moment of inertia $I=25.4$ u\AA$^2$ in the case of $\beta-$BGS, where the kinetic energy is scaled by $E_K=\hbar^2/(2I)=E_K=0.96$ K.
Thus, we have the situation $E_K < V_h$.

Here, we mention that $\beta$-BGG are slightly off-centered by 0.15 \AA~\cite{18christ, 18jiang} with much smaller dipole moments compared with the case of $\beta$-BGS. In this connection, it is noteworthy that $p$-type $\beta$-BGG shows crystalline-like thermal conductivity, while $n$-type $\beta$-BGS behaves glass-like~\cite{02avila1}.
This is one of the evidences that the long-range dipole-interaction is crucial for interpreting the thermal properties of type-I clathrates since the Coulomb interaction between the cages and guest cations Ba$^{2+}$ in $\beta$-BGG is shielded in $n$-type electron-rich BGG, while the dipole interaction in $p$-type becomes relevant even for small dipole moments.
The network configuration consisting of the off-centered guest ions is schematically illustrated in Fig.~1.

\begin{figure}[htbp]
\begin{center}
\includegraphics[width = 0.9\linewidth]{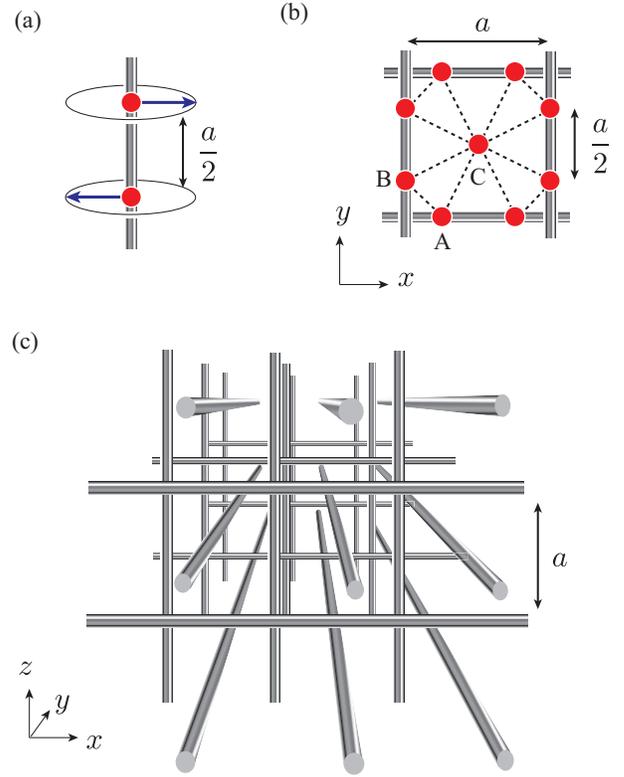}
\caption{(Color online) Schematic illustration of the configuration of the guest ions in the over-sized cages in $\beta$-BGS.
The four-fold inversion axes are directed along $x$, $y$, $z$.
(a) The deviation from the on-center position (filled circles) induces the electric dipoles (arrows).  Here the two nearest dipoles are depicted.  The electric dipoles rotate in the plane perpendicular to the axis linking the nearest dipoles.
(b) The configuration of the on-center positions of the guest ions.  The filled circles represent the positions of off-centered guest ions, around which the electric dipoles are induced.  The sites A, B, and C in (b) are seated on the chains parallel to $x$, $y$, and $z$, respectively: $A=(a/4,0,a/4)$, $B=(0,a/4,3a/4)$, $C=(a/2,a/2,a/2)$.  The distance between the next-nearest neighbors (dashed lines) is $\sqrt{3/8}\,a$.
Note that these three dipoles constitute an \emph{equilateral triangle} and easily generate a frustrated situation.
(c) The 3D configuration of the dipole chains is illustrated.
} \label{fig:config}
\end{center}
\end{figure}

The separation of neighboring wells is small, for example, $\pi r_0/2= 0.67$ {\AA} in $\beta$-BGS.
This allows the guest ion to tunnel to a nearby potential minimum at lower temperatures than the barrier height energy $V_h$, where the off-centered guest ion executes zero-point motion at one of the four wells with an energy $\hbar \omega_0$.
We can take into account only two wells with a small energy difference $\epsilon=E_1-E_2$ (the asymmetric energy of neighboring double wells),
which mainly attribute to the disordered configuration of the Ga$^{-}$ and Sn framework atoms of the cages.

In this case, the Hamiltonian is reduced to a 2$\times$2 matrix with the off-diagonal element $\Delta$ (the tunneling term due to the overlap of the wavefunctions), and the energies are given by $E=\pm(1/2) \sqrt{\epsilon^2+\Delta^2}$.
The specific heat incorporating all of the off-centered guest ions is given by
\begin{eqnarray}
C = 2k_B^2 T \int_{0}^{\infty} x^2 \sech^2(x) P(2x k_B T)dx,
\label{C}
\end{eqnarray}
where $x$ is the dimensionless variable defined by $E/(2k_BT)$, and $P(E)$ is the number density of states per volume;
namely, $P(E)dE$ is the number of tunneling states per volume with energy between $E$ and $E+dE$ as described in~\cite{19anderson,20phillips}.

\section{Noninteracting-dipole picture}
The picture that tunneling between nearby sites in $V_h(\theta)$ is relevant to the glass-like thermal properties often leads to the erroneous idea that the \textit{noninteracting}-dipole picture would work to explain the glass-like behaviors.
Such a model, however, does \textit{not} reproduce the observed temperature dependence and the magnitudes of the specific heats for type-I clathrate compounds with off-centered guest ions as shown below.
In fact, as will be discussed in the next section, the tunneling between the nearby potential minima in the configuration space generated by interacting dipoles is crucial for interpreting  glass-like thermal properties below about 1~K .

To disconfirm the \textit{noninteracting}-dipole picture, we calculate the specific heat within this picture.
When the dipole interaction between the guest-ions is irrelevant, the problem reduces to the state of only one cage.
In this case, the zero-point energy of the guest ion trapped in one of the four local-potential minima of the hindering potential $V_h(\theta)$ is estimated to be $\hbar\omega_0 \approx 4.6$ K from the uncertainty principle, taking an actual  size of box $\simeq(0.3~\mathrm{\AA})^3$ for $\beta$-BGS~\cite{18madsen}.
The tunneling energy is given by $\Delta=\hbar\omega_0\,\exp(-\sqrt{2IV_h}\delta\theta/\hbar)$ with $\delta\theta$ the angle between two nearby minima, and $\delta\theta$ takes a value $\approx\pi/2$.
If the asymmetry energy of neighboring wells $\epsilon$ is negligibly small compared with $\Delta$, we can estimate for the \textit{noninteracting}-dipole picture the most probable lower-bound energy as $\Delta_{min}=0.03$ K by using values for $\beta$-BGS:
$\hbar\omega_0=4.6$ K and $\delta\theta=\pi/2$.
Note that, in the case of the \textit{interacting}-dipole picture, $\Delta_{min}$ becomes much smaller (see below).

The lower and the upper bounds of the integral in Eq.~(\ref{C}) should be $\Delta_{min}/(2k_BT)$ and $E_{max}/(2k_BT)$, respectively, depending on the temperature $T$.
Since the function $x^2\sech^2(x)$ in the integrand has a maximum at around unity, the contribution to the integral of the product of $x^2\sech^2(x)$
and $P(2xk_BT)$ becomes small at temperatures below $T \leq \Delta_{min}$.
In addition, the \textit{noninteracting}-dipole picture claims that \textit{every} off-centered guest ion contributes to the specific heat.
By taking the width $\bar{E}$ of the $P(E)$ from $\Delta_{min}$ to $ E_{max}\approx 2V_h$ and by incorporating the number of guest ions in a unit cell 6, Eq.~($\ref{C}$) yields $C\simeq4100$ $\mathrm{mJ\,mol^{-1}\,K^{-1}}$ at 1 K, which is two orders larger than the observed value for $\beta$-BGS $C\simeq30$ $\mathrm{mJ\,mol^{-1}\,K^{-1}}$ at 1~K~\cite{02avila1,04avila2,05suekuni2}.
Thus, the \textit{noninteracting}-dipole picture severely conflicts with the observations.

In this connection, it is noteworthy that the Einstein model has been used to analyze the Boson peak-like behavior of specific heats or inelastic neutron scattering data observed at the around 10 K energy region.
As is well known, the Einstein model is a good phenomenological approximation for describing the excess phonon density of states, but is not suitable to grasp the essential physical picture behind the model such as the symmetry or many-body interaction involved in the systems.

\section{Interacting-dipole picture}
Electric dipole moments of guest ions provide long-range interactions with  dipoles in other cages.
To make our argument clearer,
let us consider at first two electric dipoles $\vec{p}_i$ and $\vec{p}_j$ separated by a distance  $|\vec{R}_{ij}|$.
The dipolar interaction gives the following form under the condition $r_0 \ll R_{ij}$:
\begin{eqnarray}
 \tilde{V}_{ij}(\vec{p}_i,\vec{p}_j)=
\frac{\vec{p}_i\cdot\vec{p}_j - 3(\vec{p}_i\cdot\hat{R}_{ij}) (\vec{p}_j\cdot\hat{R}_{ij})}{4\pi\varepsilon_r|\vec{R}_{ij}|^3},
\label{Vij}
\end{eqnarray}
where $\varepsilon_r$ is the dielectric constant of the clathrate considered here, and $\hat{R}_{ij}=\vec{R}_{ij}/|\vec{R}_{ij}|$.
In the case of the $\beta$-BGS clathrate, off-centered guest ions are in 14-hedron cages, and its nearest-neighbor off-centered guest ions are located in the next two  14-hedron cages, which share the \textit{same} four-fold inversion axis.
Figure 1 gives the configuration of the guest ions in over-sized cages.
The four-fold inversion axes are directed along the $x, y, z$ axes due to the cubic symmetry of $\beta$-BGS.

The potential function for the two coupled dipoles $\vec{p}_1$ and $\vec{p}_2$
becomes $V_{12}=V_h(\theta_1)+V_h(\theta_2)+W_{12}(\theta_1, \theta_2)$
with $W_{12}=p^2\cos(\theta_1-\theta_2)/(4\pi\varepsilon_r R_{12}^3)$,
where two global minima (maxima) in $(\theta_1,\theta_2)$ configuration space appear at  $|\theta_1-\theta_2|=\pi$~$( 2\pi)$
since the dominant term for a nearest-neighbor pair is the first term in Eq.~(\ref{Vij}).
This configuration acts as a new hindering potential in addition to the four-fold inversion symmetric potential $V_h$.
This argument can be straightforwardly extended to the case of multiple pairs providing many potential minima in the configuration space $\mathcal{P}=(\theta_1,\theta_2,\theta_3, \cdots)$~\cite{21shima}, where the potential function is $V_{123\cdots}= \sum V_h(\theta_i)+\sum\tilde{V}_{ij}$.

The energy scale of the dipolar interactions between nearest neighbors is given by its maximum value:
\begin{eqnarray}
J_1=\frac{p^2}{4\pi\varepsilon_rR_1^3}
\label{J}
\end{eqnarray}
where $R_1$ is the distance between the nearest neighbors.
We also represent the energy scale of the next-nearest neighbor coupling as $J_2\simeq J_1/2$.
Here, we defined $J_2$ as the maximum value of the coupling energy averaged over the eight next-nearest neighbors.
The contribution to $J_2$ from the first term in Eq.~(\ref{Vij}) is smaller than that of the second one, and mostly works as a counterforce.
Note that the motion of the dipoles is restricted in the plane perpendicular to the nearest-neighbor chain.
Therefore, we first estimate $J_2$ given by $J_2=3J_1(R_1/R_2)^3\langle\zeta\rangle_{max}$ without the first term of Eq.~(\ref{Vij}), and then take it into account, where $R_2$ ($=\sqrt{3/8}\,a$) is the distance between the next-nearest neighbors, and $\langle\zeta\rangle_{max}$ the maximum average of $\zeta=\cos(\theta_1)\cos(\theta_2)$ over the angles.
Setting $\theta_2$ as a function of $\theta_1$ which yields the largest $\zeta$ for each $\theta_1$, we average $\zeta$ over $\theta_1$.
Since $\theta_1$ takes a value between $\theta_{min}$ and $(\pi-\theta_{min})$, where $\theta_{min}=\tan^{-1}(1/\sqrt{5})$ in $\beta$-BGS, it follows $\langle\zeta\rangle_{max}\simeq0.5$ and $J_2\simeq 0.8J_1$.
 The average of the first term over eight neighbors becomes $\simeq-0.3J_1$ in the case that four next-nearest dipoles are anti-parallel and the others are perpendicular to the central dipole.
Since the contribution to $J_2$ from the first term is not far from this, we can take $J_2\simeq 0.5J_1$.

The actual distance between nearest-neighbor guest ions in $\beta$-BGS is $R_{1}=a/2=5.84$ {\AA}, then the characteristic energy scale for nearest-neighbor coupling is estimated as $J_1\simeq 6\varepsilon_r/\varepsilon_0$ K.
Taking into account the dielectric constant for semiconductors in the range $5\leq\varepsilon_r/\varepsilon_0\leq20$, it turns out that the guest ions experience strong interactions.
They are no longer regarded as independent dipoles.
Combining four-fold inversion symmetry of dipoles with the frustrated situation due to the equilateral triangle (See Fig.~1), many local minima are created in a hierarchical potential map in a configuration space $\mathcal{P}$, where tunneling should involve simultaneous local structural rearrangements of an appropriate number of guest ions.

An EXAFS study \cite{18christ} for $\beta$-BGG has shown that the off-center sites actually have a three dimensional (3D) components.
Though our analysis is based on the 2D picture, our main conclusions are not altered even taking account of 3D character since the key aspect is to generate multi-valley potentials in the configuration space by long-range dipole-interaction.

$P(E)$ should be symmetric about $E=0$ since $E_1>E_2$ or $E_1<E_2$ is equally likely, implying $P(E)$ is an even function of $E$.
It is reasonable, at the first glance, to take $P(E)$ as a  normal distribution at the center of the distribution.
We employ the form of $P(E)$ as
\begin{eqnarray}
P(E)=\frac{2n \,V}{\bar{E}\sqrt{\pi}}\exp\left[-(E/\bar{E})^2\right],
\label{P}
\end{eqnarray}
where $V=N_A a^3$ ($N_A$:  Avogadro number), and $n$ is the number of relevant tunneling states per volume, which we set to $n=6\eta/a^3$ by incorporating the number of off-centered guest ions, 6, in a unit cell $a^3$.
The numerical factor $\eta$ expresses how many of the off-centered guest ions are relevant to the specific heat in effect: Note that $1/\eta$ is the averaged number of off-centered guest ions involved in a rearrangement of the configuration.

\section{Explicit form of $T$-linear specific heat}
By taking the upper bound ($\simeq 2\bar{E}$) from Eq.~(\ref{P}) as the maximum coupling strength from  neighbor dipoles $2\bar{E}\simeq z_1J_1+z_2J_2\simeq 6J_1$ with configuration numbers $z_1=2$ and $z_2=8$ ($J_3\ll J_1, J_2$), and noting $n/J_1=3\pi\varepsilon_r\eta/p^2$ from $R_1=a/2$,  Eq.~(\ref{C}) combined with Eq.~(\ref{P}) yields the simple form
 \begin{eqnarray}
\frac{C}{V}=\frac{\pi^{5/2} \varepsilon_r\eta}{3p^2}k_B^2T,
\label{C3}
\end{eqnarray}
for $k_BT>\Delta_{min}$.
Equation~(\ref{C3}) predicts larger specific heats for  smaller dipole moments $p$.
Suekuni \textit{et al.}~\cite{05suekuni2} have observed low-temperature specific heats  for $n$- and $p$-type $\beta$-BGS with different length dipoles of $r_0=0.434$ and 0.439 \AA, respectively.
The observed specific heats below 1 K of the $n$-type $\beta$-BGS are a few percent larger than those of the $p$-type $\beta$-BGS.
This accords with the prediction of Eq.~(\ref{C3}).

The comparison of Eq.~(\ref{C3}) with the observed magnitude of a specific heat $C\simeq 30T$ $\mathrm{mJ\,mol^{-1}\,K^{-1}}$ for $\beta$-BGS~~\cite{02avila1,03suekuni1,04avila2,05suekuni2} below 1~K~~\cite{02avila1,03suekuni1,04avila2,05suekuni2,06suekuni3} provides important information for tunneling states through the value of $\eta$.
Taking $\varepsilon_r\simeq 10\varepsilon_0$, $\eta$ is estimated to be $\simeq 0.06$ from $C\simeq 50(\varepsilon_r/\varepsilon_0)\eta T$ $\mathrm{mJ\,mol^{-1}\,K^{-1}}$.
This indicates that not all but only 6\% of the off-centered guest ions contribute to the specific heats \textit{on average}.
This implies that the averaged number of dipoles simultaneously rearranged is $1/\eta\simeq 20$, which indicates that $\Delta_{min}$ is negligible in the temperature range we consider here.

The glass transition temperature $T_g$ should be close to
$T_g\approx 2T_{melt}/3\approx 500$ K from the Tammann's rule, where $T_{melt}\approx4\bar{E}$($\sim700$ K) estimated from the effective width of the distribution function Eq.~(\ref{P}).
This temperature is much lower than the one where the dipole moments disappear in $\beta$-EGG ($>1000$ K)~\cite{07takasu1}.
We suggest an experimental study to confirm the existence of glass transition in this temperature range.

\section{Summary}
The results of our investigation highlight the role of the off-centered guest ions in $\beta$-BGS, providing a clear picture of why this compound shows glass-like behaviors.
The off-centered guest ions constitute tunneling states in a configuration space of interacting dipoles with hierarchial structure, where tunneling occurs as a rearrangement of the dipole configuration.
The rearrangement is induced in units of 5-20 dipoles.
Although these results are obtained on $\beta$-BGS, the investigations are directly applicable to other types of clathrate compounds with off-centered guest ions.

We have not discussed about the glass-like behaviors of thermal conductivities observed in type-I clathrates on the basis of our picture.
This subject will be treated in detail in our forthcoming paper.

\acknowledgments
This work was supported by the U.S. DOE, Office of Science, Office of Basic Energy Sciences, under Contract No. DE-FG02-05ER46241 at MIT, and the U.S. DOE, Office of Science, Office of Basic Energy Sciences, under Contract No. DE-AC02-06CH11357 at Argonne National Laboratory.

\end{document}